\documentclass[times,10pt,twocolumn]{article}
\usepackage{latex8}
\usepackage{times}

\input amssym.def
\input amssym
\input epsf.sty

\newtheorem{theorem}{Theorem}[section]
\newtheorem{lemma}[theorem]{Lemma}
\newtheorem{fact}[theorem]{Fact}

\newtheorem{claim}[theorem]{Claim}

\newtheorem{definition}{Definition}[section]

\newenvironment{proof}{{\bf Proof:  }}{\qquad\rule{2mm}{2mm}}
\newenvironment{proofof}[1]{{\bf Proof of #1:  }}{\qquad\rule{2mm}{2mm}}

\newcommand\meet\wedge
\newcommand\implies\Rightarrow

\def\complex{{\mathchoice {\setbox0=\hbox{$\displaystyle\rm C$}\hbox{\hbox
to0pt{\kern0.4\wd0\vrule height0.9\ht0\hss}\box0}}
{\setbox0=\hbox{$\textstyle\rm C$}\hbox{\hbox
to0pt{\kern0.4\wd0\vrule height0.9\ht0\hss}\box0}}
{\setbox0=\hbox{$\scriptstyle\rm C$}\hbox{\hbox
to0pt{\kern0.4\wd0\vrule height0.9\ht0\hss}\box0}}
{\setbox0=\hbox{$\scriptscriptstyle\rm C$}\hbox{\hbox
to0pt{\kern0.4\wd0\vrule height0.9\ht0\hss}\box0}}}}

\newcommand{\reals}{{\hbox{\sf I\kern-.14em\hbox{R}}}}
\newcommand{\trace}{{\rm Tr}}
\newcommand{\prob}[1]{{\rm Pr}\left[#1\right]}
\newcommand{\expct}[1]{{\mathrm E}\left[#1\right]}
\newcommand{\size}[1]{\left|#1\right|}

\newcommand{\ket}[1]{\left|#1\right\rangle}
\newcommand{\bra}[1]{\left\langle #1\right|}
\newcommand{\braket}[2]{\left.\left\langle #1\right|#2\right\rangle}
\newcommand{\ketbra}[2]{\ket{#1}\!\bra{#2}}
\newcommand{\norm}[1]{\left\|\,#1\,\right\|}
\newcommand{\set}[1]{{\left\{#1\right\}}}
\newcommand{\st}{{\; | \;}}

\newcommand{\ignore}[1]{}

\newcommand{\brho}{\mbox{\boldmath$\rho$}}
\newcommand{\bsigma}{\mbox{\boldmath$\sigma$}}
\newcommand{\aitch}{{\mathcal H}}

\newcommand{\oh}{{\mathcal O}}
\newcommand{\dee}{{\mathcal D}}
\newcommand{\you}{{\mathcal U}}
\newcommand{\cents}{{\rm c\!\!|\,}}
\newcommand{\pmax}{p^{\rm max}}
\title{    Optimal lower bounds for quantum automata and random access
	   codes~\thanks{This work was initiated at the 1998
	   Elsag-Bailey -- I.S.I.\ Foundation research meeting on quantum
	   computation.}}

\author{Ashwin Nayak~\thanks{
  Supported by JSEP grant FDF
  49620-97-1-0220-03-98 and NSF grant CCR 9800024.}}

\affiliation{Computer Science Division \\
             University of California \\
	     Berkeley, CA~94720}

\email{ashwin@cs.berkeley.edu}

\date{}

\begin{document}

\maketitle

\begin{abstract}
Consider the finite regular language~$L_n = \{w0 \;|\; w \in \{0,1\}^*,
|w| \le n\}$. In~\cite{antv} it was shown that while this language
is accepted by a deterministic finite automaton of size~$O(n)$, 
any one-way quantum finite automaton~(QFA) for it has
size~$2^{\Omega(n/\log n)}$. This was based on the fact
that the evolution of a QFA is required to be reversible.
When arbitrary intermediate measurements are allowed, this
intuition breaks down. Nonetheless, we show a~$2^{\Omega(n)}$
lower bound for such QFA for~$L_n$, thus also improving the previous
bound. The improved bound is obtained from
simple entropy arguments based on Holevo's theorem~\cite{holevo}.
This method also allows us to obtain an asymptotically optimal~$(1-H(p))n$ 
bound for the dense quantum codes (random access codes) introduced
in~\cite{antv}.
We then turn to Holevo's theorem, and show that in typical
situations, it may be replaced by a tighter and more transparent
in-probability bound.
\end{abstract}

\Section{Introduction}
\label{sec-intro}

One-way quantum finite automata~(QFA) were defined in~\cite{mc,kw} and
have drawn much interest since because they reflect the capabilities of
currently feasible
experimental quantum computers. Moreover, their study provides much
insight into the nature of quantum computation.
Results like those of~\cite{af} and~\cite{antv} show that
the laws underlying quantum computation are a mixed blessing.
\cite{af} shows how one may use superpositions to design QFA for certain
languages that are exponentially more succinct than the corresponding
classical FA. In contrast, other results from~\cite{antv} show that
the reversibility requirements of quantum mechanics imposes serious
limits on the power of QFA---they show that QFA for certain
other languages are exponentially larger than the corresponding DFA.
In this paper we consider a different model of QFA (called {\em enhanced\/}
QFA) where the state of the QFA can be measured while each symbol is
processed.
In the case of more general models such as quantum Turing machines
such intermediate 
measurements do not increase the power of the model, since measurements
can always be replaced by safe storage. However, in the case of QFA, 
the space limitations inherent in the definition preclude the possibility
of similar reasoning. Moreover, in this new model, the evolution of
the system is no longer reversible, so the intuition from~\cite{kw,antv}
no longer applies. Indeed, this new model of QFA was suggested by
Dorit Aharonov as a more physically appropriate model that might not
suffer from unnecessary handicaps resulting from the reversibility
property embedded in the definitions from~\cite{mc,kw}. 

In this paper, we show that enhanced QFA are also exponentially larger than
the corresponding DFA for certain languages. The conceptual framework for 
our proof is completely different from that in~\cite{antv}. We consider
the evolution of a QFA on a random input string and show that the
entropy of the mixed state that it exists in can only increase with each
successive symbol read.
This holds true even in the presence of intermediate measurements.
Moreover, for certain languages, it is possible to bound from below
the increase
in entropy that results from processing each symbol, by appealing to Holevo's
theorem~\cite{holevo}. Finally, we can bound the total information
capacity of the QFA in terms of the number of states of the QFA, and
therefore obtain a lower bound on the number of states required to correctly 
recognize strings of the language.
The new bound we get is tight, and therefore answers an issue left open 
in~\cite{antv}.

The paper~\cite{antv} also introduced the novel possibility
of dense quantum codes that seem to violate Holevo's bound by
exploiting the fact that in general measurements do not commute.
This raised the possibility of (for instance) parsimoniously encoding an
entire telephone directory such that any single number could be
extracted from it via a suitable measurement.
Examples of such {\em random access codes\/} were given in~\cite{antv}
that have no
classical counterparts. However, it was also shown that no more than a
logarithmic factor compression is achievable.
We can use the same conceptual framework as described above to
give a linear bound on the number of qubits required for such codes.
This bound is optimal up to an additive logarithmic term, as follows from the
classical upper bound given in~\cite{antv}. Thus, quantum encoding
offers no asymptotic advantage over classical encoding in this
scenario. This resolves an open question from~\cite{antv}.

Finally we turn our attention to Holevo's bound~\cite{holevo} itself.
Typically in quantum computation applications (though not in this paper),
Holevo's bound is applied by converting, often implicitly,
a statement about the probability of correct decoding
into a statement in terms of entropy, when a random
variable~$X$ is transmitted over a quantum channel using~$m$
quantum bits.
We give a tight bound on this decoding
probability by a direct argument which allows us to infer lower bounds 
for~$m$ without resorting to Holevo's theorem. Since the probability
bound is tight, the inferred bounds are at least as good as those
implied by Holevo's theorem.
We also provide an example where it gives a strictly better bound than
the latter. We should mention that the proof of Holevo's bound
(which is essentially equivalent to the strong subadditivity property
of von Neumann entropy) is rather involved, while the proof of
the probability bounds is quite transparent.

\Section{Summary of results}
\label{sec-results}

A QFA (as defined in~\cite{kw}) differs from a DFA in that its
state is in general a superposition of the classical (basis)
states. It starts in such a state, and when a new input symbol~$\sigma$
is seen, a corresponding unitary operator~$U_\sigma$ is applied to it.
The state is then measured to check for acceptance, rejection or
continuation. If the result of the measurement is `continue,' the next
symbol is read, otherwise the input is accepted or rejected. A QFA
recognizes a language if all the strings in it (or not in it) are accepted
(respectively, rejected) with constant probability bounded away from~$1/2$.
See Section~\ref{sec-qfa} for a more precise definition of QFA.

We start by showing an exponential lower bound for QFA.
\begin{theorem}
\label{thm-qfa}
Let~$L_n$ be the language
$$\{ w0 \st  w \in \{0,1\}^*,\; |w| \le n\}.$$
Then,
\begin{enumerate}
\item $L_n$ is recognized by a DFA of size~$O(n)$,
\item $L_n$ is recognized by some QFA, and
\item Any QFA recognizing~$L_n$ with some constant probability
greater than~${1\over 2}$ has~$2^{\Omega(n)}$  states.
\end{enumerate}
\end{theorem}
Note that a~$2^{\Omega(n)}$ versus~$O(n)$
separation is the best possible if only finite languages (or regular
languages with sufficiently high probability of acceptance by a QFA) are
considered:
such languages are recognized by reversible (deterministic) FA
that are at most exponentially larger than the corresponding
DFA~\cite{af}.

We then consider enhanced QFA, in which instead of only applying
a unitary transformation when a new input symbol is seen, we
allow a combination of unitary operators and orthogonal measurements. 
With the introduction of irreversibility via measurements,
it may appear that such automata be at least as powerful as DFA. However, it
is not hard to verify (by applying a technique of~\cite{rabin}, also used
in~\cite{kw}) that enhanced QFA accept only regular languages.
Moreover, we show that the bound of Theorem~\ref{thm-qfa} continues to
hold.
\begin{theorem}
\label{thm-eqfa}
The statements of Theorem~\ref{thm-qfa} hold also for enhanced QFA.
\end{theorem}
It also follows from the proof of this theorem that enhanced QFA accept
only a strict subset of the regular languages.

Random access encoding was introduced in~\cite{antv} as a potentially
powerful primitive in quantum information processing.
An {\em $(n,m,p)$-random access encoding\/}
is a function~$f$
that maps~$n$-bit strings to mixed states
over~$m$ qubits such that, for every~$i \in \set{1,\ldots,
n}$, there is a measurement~$\oh_i$ with outcome~$0$ or~$1$
that has the property that for all~$x \in \set{0,1}^n$,
$$
\prob{\oh_i(f(x)) = x_i }  \;\;\ge\;\;   p.
$$
{\em Serial encoding\/} was defined similarly, except that the
measurement~$\oh_i$ is allowed to depend on all the subsequent
bits~$x_{i+1}\cdots x_n$ of the encoded string.
The technique used in proving Theorem~\ref{thm-qfa} also yields
a bound for such encoding.
This bound matches the classical upper bound of~$(1-H(p))n + O(\log n)$ shown
in~\cite{antv} up to the logarithmic additive term.
\begin{theorem}
\label{thm-rac}
Any $(n,m,p)$-random access (or serial) encoding has~$m \ge (1-H(p))n$.
\end{theorem}

To finish, we present a simple alternative to Holevo's
bound~\cite{holevo}. 
\begin{theorem}
\label{thm-alt-holevo}
Let~$X$ be a random variable over bit strings
which are encoded as mixed states over~$m$ qubits and
let~$P(X,d)$ denote the net probability of the~$d$
most likely strings in the sample space of~$X$. 
If~$Y$ is any random variable obtained by making some measurement
of the encoding of~$X$, then 
\begin{enumerate}
\item there is a decoding procedure~$\dee_0$ such
that
$$\Pr[\dee_0(Y) = X] \;\;\ge\;\; 2^{-H(X|Y)},$$
where~$H(X|Y)$ is the
conditional Shannon entropy of~$X$ with respect to~$Y$; and
\item for any decoding function~$\dee$,
$$\Pr[\dee(Y) = X] \;\;\le\;\; P(X,2^m).$$
\end{enumerate}
\end{theorem}
In particular, this implies that when~$X$ is distributed uniformly,
the mutual information~$I(X\!:\!Y)$ of~$X$
and~$Y$ is at most~$m$. Typical applications of the Holevo's bound such
as that in~\cite{kremer,antv}
involve only this weaker form. Our bound thus obviates the need for a
translation of in-probability statements into statements about mutual
information in these cases, also giving better bounds than Holevo's theorem
in the process.

\Section{Preliminaries}

First, in Section~\ref{sec-prelims}, we review the basic elements of quantum
information theory. Then, in Section~\ref{sec-qfa},
we define enhanced QFA formally using some of the concepts
presented there.

\SubSection{Information theory basics}
\label{sec-prelims}

We use the following notation in this paper. Let~$X$ and~$Y$ be two
random variables. $H(X)$ denotes the {\em Shannon entropy\/} of~$X$;
$H(X|Y)$, the {\em conditional\/} Shannon entropy of~$X$
with respect to the variable~$Y$; and~$I(X\!:\!Y)$, the {\em
mutual information\/} of the two variables~$X$ and~$Y$. 
We also use~$H : [0,1] \rightarrow [0,1]$ to denote the binary
entropy function.  We refer the reader
to~\cite{ct} for the definition and properties of these standard
concepts from classical information theory.

The quantum mechanical analogue of a random variable is
a probability distribution over superpositions, also
called a {\em mixed state}.
Consider the mixed state~$\{p_i, \ket{\phi_i}\}$, where the
superposition~$\ket{\phi_i}$ is drawn with probability~$p_i$. The
behaviour of this mixed state is completely characterized by its {\em
density matrix\/}~$\brho = \sum_i p_i \ketbra{\phi_i}{\phi_i}$.
We will therefore identify a  mixed state with its density matrix.

The following properties of density matrices are immediate from the
definition. For any density matrix~$\brho$,
\begin{enumerate}
\item $\brho$ is Hermitian, i.e., $\brho = \brho^\dagger$.
\item $\brho$ has unit trace, i.e., $\trace(\brho) = \sum_i \brho(i,i) =
      1$.
\item $\brho$ is positive semi-definite, i.e.,
      $\bra{\psi}\brho\ket{\psi} \ge 0$ for all~$\ket{\psi}$.
\end{enumerate}
Thus, every density matrix is {\em unitarily diagonalizable\/} and has
non-negative real eigenvalues that sum up to~$1$. 
The {\em von Neumann entropy\/}~$S(\brho)$ of a density
matrix~$\brho$ is defined as~$S(\brho) = - \sum_i \lambda_i \log \lambda_i$,
where~$\{\lambda_i\}$ is the multiset of all the eigenvalues
of~$\brho$. In other words, $S(\brho)$ is the Shannon entropy of the
distribution induced by the eigenvalues of~$\brho$ on the corresponding
eigenvectors.
For a comprehensive introduction to this concept and
its properties, see, for instance,~\cite{wehrl, peres, preskill}.

The density matrix corresponding to a mixed state with superpositions
drawn from a Hilbert space~$\aitch$ is said to have {\em support\/}
in~$\aitch$. First, we note the following.

\begin{fact}
\label{fac-entropy-ub}
If~$\brho$ is a density matrix with support in a Hilbert space of
dimension~$d$, then~$S(\brho) \le \log d$.
\end{fact}
This is because the probability distribution induced by the
eigenvalues of~$\brho$ has support of size at most~$d$. The
Shannon entropy
of any such distribution is at most~$\log d$.

When a unitary operator~$U$ is
applied to a mixed state, the corresponding density matrix~$\brho$ is
transformed to~$U\!\brho U^\dagger$. Since the eigenvalues of~$U\!\brho
U^\dagger$ are the same as those of~$\brho$, 
we conclude that entropy is invariant under unitary operations.

\begin{fact}
\label{fac-invariant}
For any density matrix~$\brho$ and unitary operator~$U$, we
have~$S(U\!\brho U^\dagger) = S(\brho)$.
\end{fact}
On the other hand, when we make an orthogonal measurement 
on a mixed state, the the entropy of the system can only
increase.\footnote{ This fact
may appear to be counterintuitive at first, since the entropy of a
system is usually understood to quantify our ignorance of the
state of the system, and making a measurement reveals some information
about its state. However, it should be noted that the increase in
entropy is not claimed in the state of the system conditioned on
the state of the observer, but in the state of the system with the
state of the observer traced out.}
If a mixed state~$\brho$ is measured according to
an orthogonal set of projections~$\{ P_j \}$,
it is easily verified that the resulting
density matrix is given by~$\sum_j P_j\brho P_j$.

\begin{fact}
\label{fac-monotone}
Let~$\brho$ be the density matrix of a mixed state in a Hilbert
space~$\aitch$ and let the set of orthogonal projections~$\{ P_j \}$
define a measurement in~$\aitch$.
Further, let~$\rho' = \sum_j P_j\brho P_j$ be the density matrix resulting
from a measurement of the mixed state with respect to this observable.
Then~$S(\brho') \ge S(\brho)$.
\end{fact}
It is not hard to see that this is in fact a consequence of
the property of density
matrices that the entropy of any random variable obtained by making a
measurement on  a
mixed state is at least as much as the entropy of its density 
matrix. 
A proof of this property may be found in~\cite[Chapter~9,
pp.~262--263]{peres}.

\SubSection{Enhanced one-way quantum finite automata}
\label{sec-qfa}

An enhanced one-way quantum finite automaton~(QFA) is a theoretical
model for a quantum computer with finite workspace. Models for such
space-restricted quantum computers were first considered by~\cite{mc,kw}.
However, these models did not include the full range of operations 
allowed by the laws of quantum mechanics. In particular, the model
of~\cite{mc} does not include measurements as intermediate steps in
a computation, and the model of~\cite{kw} allows only measurements that
check for acceptance, rejection or continuation. The model we describe
below rectifies this situation by allowing any {\em orthogonal\/}
measurement as a valid intermediate computational step.
Our model may be seen as a finite memory
version of the mixed state quantum computers defined in~\cite{akn}.
Note that we do not allow the more
general ``positive operator valued measurements'' because the
implementation of such measurements involves the joint unitary evolution
of the state of the automaton with a fresh
set of ancilla qubits, which runs against the (fixed finite workspace)
spirit of the model.

In abstract terms, we may define an enhanced QFA as follows.
It has a finite set of basis states~$Q$, which consists of three parts:
accepting states, rejecting states and non-halting states.
The sets of accepting, rejecting and non-halting basis states are
denoted by~$Q_{\rm acc}, Q_{\rm rej}$ and $Q_{\rm non}$, respectively.
One of the states,~$q_0$, is distinguished as the starting state.

Inputs to a QFA are words over a finite alphabet~$\Sigma$.
We shall also use the symbols~`$\cents$' and~`$\$$'
that do not belong to~$\Sigma$ to 
denote the left and the right end-marker, respectively.
The set~$\Gamma = \Sigma\cup\{\cents, \$ \}$ denotes the working alphabet
of the QFA.
For each symbol~$\sigma\in\Gamma$, an enhanced QFA has a corresponding
``superoperator''~$\you_{\sigma}$ which is given by a composition of a
finite sequence of unitary transformations and orthogonal measurements 
on the space~$\complex^Q$.
An enhanced QFA is thus defined by describing~$Q, Q_{\rm acc},
Q_{\rm rej}, Q_{\rm non}, q_0, \Sigma$, and~$\you_\sigma$ for
all~$\sigma\in\Gamma$.

At any time, the state of a QFA can be described by a density matrix
with support in~$\complex^Q$. The computation starts in the 
state~$\ketbra{q_0}{q_0}$. Then transformations corresponding to 
the left end marker~`$\cents$,' the letters of the input word~$x$ and
the right end marker~`$\$$' are applied in succession to the state of
the automaton, unless a transformation results in acceptance or rejection
of the input. A transformation corresponding to a symbol~$\sigma\in\Gamma$
consists of two steps:
\begin{enumerate}
\item
First, $\you_\sigma$ is applied to~$\brho$, the current state of the
automaton, to obtain the new state~$\brho'$.
\item
Then, $\brho'$ is measured with respect to 
the observable~$E_{\rm acc}\oplus E_{\rm rej}\oplus E_{\rm non}$,
where $E_{\rm acc}={\rm span}\{|q\rangle \;|\; q\in Q_{\rm acc}\}$,
$E_{\rm rej}={\rm span}\{|q\rangle \;|\; q\in Q_{\rm rej}\}$,
$E_{\rm non}={\rm span}\{|q\rangle \;|\; q\in Q_{\rm non}\}$.
The probability of observing~$E_i$ is
equal to~$\trace(P_i\brho')$, where~$P_i$ is the orthogonal projection
onto~$E_i$.
If we observe~$E_{\rm acc}$ (or~$E_{\rm rej}$), the input is accepted
(or rejected). Otherwise, the computation continues (with the
state~$P_{\rm non}\brho'P_{\rm non}$), and the 
next transformation, if any, is applied.
\end{enumerate}
We regard these two steps together as reading the symbol~$\sigma$.

A QFA~$M$ is said to {\em accept\/} (or {\em recognize}) a language~$L$ with
probability~$p > {1\over 2}$ if 
it accepts every word in~$L$ with probability at least~$p$, 
and rejects every word not in~$L$ with probability at least~$p$.

The {\em size\/} of a finite automaton is defined
as the number of (basis) states in it. The ``space used by the
automaton'' refers to the number of (qu)bits required to represent an
arbitrary automaton state.

The model of QFA as defined in~\cite{kw} differs from this model in that
the superoperators~$\you_\sigma$ are all required to be given by unitary
transformations~$U_\sigma$.

\Section{The automata and coding lower bounds}

In this section, we prove the first three theorems of
Section~\ref{sec-results}. They are all based on a common framework
which we present in Section~\ref{sec-key}. 

\SubSection{The conceptual framework}
\label{sec-key}

\begin{figure}
\label{fig-entropy}
\begin{center}
\epsfxsize=3.2in
\hspace{0in}
\epsfbox{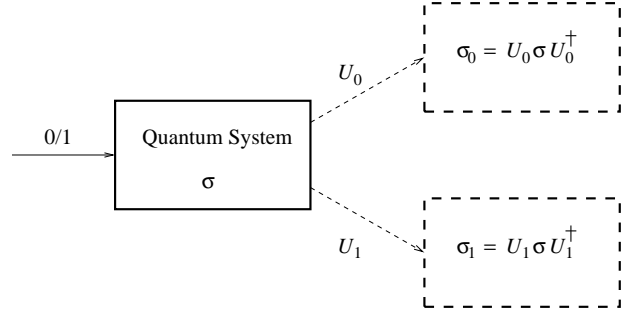}
\caption{\it A stream of random bits determining the evolution of a
quantum system.}
\end{center}
\end{figure}

Consider the evolution of the a quantum system under a random
sequence of
unitary transformations~$(V_i)$, where each~$V_i$ is
either~$U_0$ or~$U_1$
(see Figure~1).
Now suppose that the transformations~$U_0$ and~$U_1$
are distinguishable in the sense that for every superposition~$\ket{\phi}$
of the system, $U_0\ket{\phi}$ can be distinguished
from~$U_1\ket{\phi}$ with success probability, say, $2/3$ by some
fixed measurement. At each step, the system gains
some information about the transformation applied to it, and we expect
the entropy of the system to increase accordingly. In general, we could
apply one of two arbitrary but distinguishable quantum operations
on the system, and we would expect the same increase in entropy.
This is the essential content of our key lemma:

\begin{lemma}
\label{lem-mix}
Let~$\bsigma_0$ and~$\bsigma_1$ be two density matrices, and
let~$\bsigma = {1\over 2}(\bsigma_0 + \bsigma_1)$ be a
random mixture of these matrices.
If~$\oh$ is a measurement with outcome~$0$
or~$1$ such that making the measurement on~$\bsigma_b$
yields the bit~$b$ with average probability~$p$, then
\begin{eqnarray*}
S(\bsigma) & \ge & {1\over 2}[S(\bsigma_0) + S(\bsigma_1)]
                   + (1-H(p)).
\end{eqnarray*}
\end{lemma}
This lemma is a simple corollary of the classic Holevo
theorem~\cite{holevo} from quantum information theory which bounds
the amount of information we can extract from a quantum encoding
of classical bits.

\begin{theorem}[Holevo]
\label{thm-holevo}
Let~$x \mapsto \bsigma_x$ be any quantum encoding of bit strings,
let~$X$ be a random variable with a distribution given by~$\prob{X=x} =
p_x$, and let~$\bsigma = \sum_x p_x \bsigma_x$ be the state corresponding to
the encoding of the random variable~$X$. If~$Y$ is any random variable
obtained by performing a measurement on the encoding, then
\begin{eqnarray*}
I(X\!:\!Y) & \le & S(\bsigma) - \sum_x p_x S(\bsigma_x).
\end{eqnarray*}
\end{theorem}

\noindent
\begin{proofof}{Lemma~\ref{lem-mix}}
Consider~$\bsigma_b$ to be an encoding of the bit~$b$. 
If~$X$ is an unbiased boolean random variable, then~$\bsigma$ represents
the encoding of~$X$. Let~$Y$ be the outcome of the
measurement of this encoding according to~$\oh$.
By the hypothesis of the lemma, $\prob{Y = X} = p$. It is easy to see
from the concavity of the entropy function that
$$I(X\!:\!Y) \;\;\ge\;\; 1 - H(p)$$
(cf.\ Fano's inequality~\cite{ct}).
The lemma now follows from Theorem~\ref{thm-holevo}.
\end{proofof}

\SubSection{The case of quantum automata}

We now prove Theorem~\ref{thm-qfa} using this framework. The first two
parts of the theorem are easy; we turn to part~3. We need the following
definition from~\cite{antv}. 

\begin{definition}
An {\em $r$-restricted\/} one-way QFA for a language~$L$
is a one-way QFA that recognizes the language with probability~$p >
{1\over 2}$, and which halts with non-zero probability before
seeing the right end-marker
only after it has read~$r$ letters of the input.
\end{definition}

We first prove a bound of~$2^{(1-H(p))n}$ for the number of basis 
states in
any $n$-restricted QFA~$M$ for~$L_n$. Note that the evolution of~$M$ 
on reading stream of random bits
corresponds exactly to that of the quantum system described in
Section~\ref{sec-key} during the first~$n$ steps.
So, at the end of reading a random $n$-bit
string, the state of~$M$ has entropy at least~$(1-H(p))n$. However, this
entropy is bounded by~$\log\size{Q}$ by Fact~\ref{fac-entropy-ub} above,
where~$Q$ is the set of basis states of~$M$. This gives us the above
bound. Since we will refer to this argument later, we formalize it below.

Let~$\brho_k$ be the state of the QFA~$M$ after the~$k$th
symbol of a random $n$-bit input has been read ($0\le k \le n$).

\begin{claim}
\label{claim-incr}
$S(\brho_k) \ge (1-H(p))k$.
\end{claim}
\begin{proof}
Let~$U_\sigma$ be the unitary operator of~$M$ corresponding to the
symbol~$\sigma$. Let~$E_0$
be the span of the accepting basis states of~$M$ and let~$E_1$ be the
subspace orthogonal to it. Define the measurement~$\oh$
as applying the transformation~$U_\$$ (recall that~`\$' is the right
end-marker) and then measuring with respect to
the observable~$E_0 \oplus E_1$. We can now prove the claim by induction.

For~$k = 0$, the state of the automaton is pure, so~$S(\brho_0) = 0$.
Now assume that~$S(\brho_{k-1}) \ge (1-H(p))(k-1)$. After the~$k$th
random input symbol is read, the state of~$M$ becomes
$$\brho_k \;\;=\;\; {1\over 2}(U_0\brho_{k-1}U_0^\dagger
+ U_1\brho_{k-1}U_1^\dagger).$$
By the definition of~$M$, measuring~$U_b\brho_{k-1}U_b^\dagger$ according
to~$\oh$ yields~$b$ with probability at least~$p > {1\over 2}$.
So by Lemma~\ref{lem-mix}, we have
\begin{eqnarray}
\label{eqn-star}
S(\brho_k) & \ge & {1\over 2}\sum_{b = 0,1} S(U_b\brho_{k-1}U_b^\dagger)
              + (1-H(p)).
\end{eqnarray}
But the entropy of a mixed state is preserved by unitary transformations
(Fact~\ref{fac-invariant}),
so
$$ S(U_b\brho_{k-1}U_b^\dagger) \;\;=\;\; S(\brho_{k-1}) \ge (1-H(p))(k-1).$$
Inequality~(\ref{eqn-star}) now gives us the claimed bound.
\end{proof}

To pass from a bound on restricted QFA to one for general
QFA for the language, we now invoke the following lemma
from~\cite{antv}.
\begin{lemma}
\label{lem-restr}
Let~$M$ be a one-way QFA with~$\size{Q}$ states
recognizing a language~$L$ with probability~$p$.
Then there is an~$r$-restricted one-way QFA~$M'$ with~$O(r\size{Q})$ states
that recognizes~$L$ with probability~$p$.
\end{lemma}
Thus, any general QFA for~$L_n$ using~$\size{Q}$ basis states yields
an $n$-restricted QFA that uses~$O(n\size{Q})$ states.
By the lower bound derived above, we then have
$$ \size{Q} \;\;\ge\;\; 2^{(1-H(p))n - \log n - O(1)},$$
the bound stated in Theorem~\ref{thm-qfa}.

\SubSection{Robustness of the automata lower bound}

As mentioned in Section~\ref{sec-intro}, QFA in which general
intermediate measurements are allowed (which we call enhanced QFA),
were suggested as a way of
overcoming the restriction of reversible evolution that leads
to the exponential lower bound shown in~\cite{antv} (and in the previous
section). Theorem~\ref{thm-eqfa} rules
out this possibility. We prove this next.

Armed with the formalism of density matrices,
it is not hard to verify (by using a technique of~\cite{rabin}, which is
also used in~\cite{kw}) that
enhanced QFA accept only regular languages. Moreover, the lower bound
of Theorem~\ref{thm-qfa} continues to hold for such QFA, as we show
below. This essentially follows from the fact that the entropy of a
quantum system {\em cannot decrease\/} under the action of a sequence of
unitary operations and orthogonal measurements.

We now sketch how the proof of Theorem~\ref{thm-eqfa} may be
completed. We proceed as in the previous section by first showing the
bound for {\em restricted\/} enhanced QFA, which are defined analogously.
Lemma~\ref{lem-restr}, which extends easily to enhanced QFA, then
gives us the claimed bound. 

As before, we consider the state of a restricted automaton for~$L_n$
with acceptance probability~$p$ after a random $n$-bit input has
been read. Its entropy is bounded
by~$\log \size{Q}$, where~$Q$ is the set of its basis states.
Following Lemma~\ref{claim-incr}, we argue that the entropy of the
automaton state increases by at least~$1-H(p)$ every time a new random
input symbol is read. Claim~\ref{claim-incr} extends easily to this case
as well: initially, $S(\brho_0) \ge 0$, and 
we need only prove that~$S(\you_b\brho_{k-1}) \ge S(\brho_{k-1})$
for~$b = 0,1$, where~$\you_b$
is the superoperator corresponding to the bit~$b$, and~$\brho_i$ is the
density matrix of the automaton state after~$i$ input symbols have been read.
Since~$\you_b$ is the composition of a finite sequence of unitary
operators and orthogonal measurements, this is immediate from the
monotonicity property of density matrices implied by Facts~\ref{fac-invariant}
and~\ref{fac-monotone}.

As a simple consequence, we obtain:
\begin{theorem}
The regular language~$\{0,1\}^*0$ cannot be accepted by any enhanced QFA
with probability bounded away from~$1\over 2$.
\end{theorem}
To see this, we note that any enhanced QFA that supposedly recognizes
this language also correctly recognizes all words of length at most~$n$ of the
language~$L_n$, for every~$n$. The proof of Theorem~\ref{thm-eqfa}
now tells us that the number of states in the QFA is~$2^{\Omega(n)}$ for
every~$n$, which is a contradiction.

\SubSection{Random access codes}
\label{sec-rac}

We now prove Theorem~\ref{thm-rac}. Consider any random access encoding
with parameters~$n,m,p$.
Let~$\brho_x$ denote the density matrix corresponding to the encoding of
the $n$-bit string~$x$.
The density matrix of a random codeword is given by~$\brho =
{1\over {2^n}} \sum_x \brho_x$. We can bound the entropy of~$\brho$
by~$m$ by Fact~\ref{fac-entropy-ub}.
Using Lemma~\ref{lem-mix}, we can also prove a lower bound for the
entropy of~$\brho$, and hence obtain a lower bound on~$m$.

For any~$y \in \set{0,1}^k$, where~$0 \le k \le n$, let
$$\brho_y \;\;=\;\; {1 \over {2^{n-k}}} \sum_{z \in \set{0,1}^{n-k}} \brho_{zy}.$$
We claim that
\begin{claim}
$S(\brho_y) \ge (1-H(p))(n-k)$.
\end{claim}
\begin{proof}
The proof is by downward induction on~$k$.
The base case~$k = n$ is satisfied
easily:~$S(\brho_y) \ge 0$ for all~$n$-bit strings~$y$.

Suppose the claim is true for~$k+1$. We have
$$\brho_y \;\;=\;\; {1\over 2}(\brho_{0y} + \brho_{1y}).$$
By hypothesis,
$$S(\brho_{by}) \;\;\ge\;\; (1-H(p))(n-k-1),$$
for~$b=0,1$. Moreover, since the two density matrices
are mixtures arising from strings that differ in the~$(n-k)$th bit, the
measurement~$\oh_{n-k}$ distinguishes them correctly with probability~$p$.
Thus, by Lemma~\ref{lem-mix}, we get
$$S(\brho_y) \;\;\ge\;\; {1\over 2}(S(\brho_{0y}) + S(\brho_{1y})) + (1-H(p)),$$
which gives us the claimed bound.
\end{proof}

Theorem~\ref{thm-rac} now follows by combining the claim (with~$y$
chosen to be the empty string) and the upper bound of~$m$ on the
entropy. Notice that we could allow the measurement~$\oh_i$ to depend on
the subsequent bits of the encoded string in the argument above.
This means that the bound holds for serial codes as well.

We conclude this section by observing that the bound of Theorem~\ref{thm-rac}
also gives a communication lower bound for the problem of
information-theoretically secure private information retrieval with one
database
(see, e.g.,~\cite{cgks}). The problem may be described as the following
communication game. One party, Alice, has as input
an~$n$-bit string~$x$ (the
database) and the second party, Bob, has an index~$i \in \{ 1,\ldots, n
\}$. Bob wishes to learn the value of the~$i$th entry in the
database~$x_i$ (with probability~$p > {1\over 2}$)
without revealing any information about~$i$ to Alice.
The privacy condition translates to the fact that in any (quantum)
protocol for this problem, Bob's computation and communication are
independent of his input. We may also assume (by the principle of safe
storage) that no intermediate measurements are made during the 
quantum protocol.
A lemma due to~\cite{kremer} (based on a technique
from~\cite{yao}) 
tells us that whenever Bob's actions in a protocol 
are oblivious to his input, his state lies in a fixed subspace of
dimension~$2^m$ independent of Alice's input, if~$m$ qubits were
exchanged during the protocol.
Since his state at the end of an information retrieval protocol
is independent of~$i$, Bob may extract {\em any\/} bit~$x_j$ from the
state by making a suitable measurement. Thus, an~$m$-qubit 
protocol defines a random
access code over~$m$ qubits, which implies that~$m \ge (1-H(p))n$.

\Section{An alternative to Holevo's theorem}
\label{sec-hol}

In this section, we prove Theorem~\ref{thm-alt-holevo}.
We first prove the lower bound on the decoding probability.

Consider random variables~$X$ and~$Y$ as in the statement of
Theorem~\ref{thm-alt-holevo}. We describe a natural 
decoding procedure~$\dee_0$ and then
show that it satisfies the requirement of the theorem. On input~$y$, the
decoding algorithm outputs~$x$ such that~$p_{x|y} = \max_{x'}\;p_{x'|y}$,
where~$p_{x|y} = \Pr[X=x|Y=y]$. Let~$\pmax_y$ denote this probability
and let~$x_y$ denote the corresponding~$x$.
\begin{claim}
The procedure~$\dee_0$ described above decodes correctly with probability at
least~$2^{-H(X|Y)}$.
\end{claim}
\begin{proof}
The probability of correct decoding is equal to
\begin{eqnarray*}
\lefteqn{ \Pr[\dee_0(Y) = X] } \\
 & = & \sum_y \Pr[X=x_y|Y=y]\cdot \Pr[Y=y] \\
 & = & \expct{\pmax_Y}.
\end{eqnarray*}
Now, $H(X|Y=y) = -\sum_x p_{x|y} \log p_{x|y} \ge -\log \pmax_y$.
So~$\pmax_y \ge 2^{-H(X|Y=y)}$. Taking expectation over~$Y$, and noting
that~$2^{-(\cdot)}$ is a convex function, we have
\begin{eqnarray*}
\expct{\pmax_Y} & \ge & \expct{2^{-H(X|Y=y)}} \\
                & \ge & 2^{-\expct{H(X|Y=y)}} \\
		& =   & 2^{-H(X|Y)},
\end{eqnarray*}
which gives us the claimed lower bound on the decoding probability.
\end{proof}

We now turn to the upper bound on the probability of correct decoding.
Consider any encoding of strings~$x \mapsto \{q_{x,i},
\ket{\phi_{x,i}}\}$ into mixed states over~$m$ qubits, and any decoding
procedure~$\dee$. The output of~$\dee$ may be viewed as the outcome of a
measurement given by orthogonal projections~$\{P_x\}$ in the Hilbert
space of the encoding augmented with some ancilla. The
probability may then be bounded as
\begin{eqnarray}
\lefteqn{ \Pr[\dee(Y) = X] } \nonumber \\
& = & \sum_x \Pr[\dee(Y) = x]\cdot\Pr[X=x] \nonumber \\
& = & \sum_x p_x \sum_i q_{x,i} \norm{P_x\ket{\phi_{x,i}}}^2 \nonumber \\
\label{eqn-starstar}
& \le & \sum_x p_x \norm{P_x\ket{\phi_x}}^2,
\end{eqnarray}
where~$ p_x = \Pr[X=x]$, and~$\ket{\phi_x}$ is the pure
state~$\ket{\phi_{x,i}}$ that maximizes the
probability~$\norm{P_x\ket{\phi_{x,i}}}^2$ of obtaining the correct
outcome~$x$ when its encoding is measured. (In all the expressions in
this section, the
ancilla qubits used in the measurement have been suppressed for ease of
notation.) We can now bound the decoding probability by using the
following claim.
\begin{claim}
$\sum_x \norm{P_x\ket{\phi_x}}^2 \le 2^m$.
\end{claim}
\begin{proof}
Let~$E$ be the subspace spanned by the codewords~$\ket{\phi_x}$, and
let~$Q$ be the projection onto~$E$. Since
the codes are over~$m$ qubits, $E$ has dimension at most~$2^m$.
Let~$\{\ket{e_i}\}$ be an orthonormal basis for~$E$.
Let~$\{\ket{\hat{e}_{x,j}}\}$ be an orthonormal basis for the range of~$P_x$.
The union of all these bases~$\{\ket{\hat{e}_{x,j}}\}$
is an orthonormal basis for the entire decoding Hilbert space.
Now,
\begin{eqnarray*}
\norm{P_x\ket{\phi_x}}^2 & = &  
\sum_j \size{\braket{\hat{e}_{x,j}}{\phi_x}}^2 \\
& \le & \sum_j \norm{Q\ket{\hat{e}_{x,j}}}^2.
\end{eqnarray*}
The last inequality follows because the length of the projection
of any vector onto a space~$W$ is at least the length of its
projection onto a subspace~$V$ of~$W$.
Observe that~$\norm{Q\ket{\hat{e}_{x,j}}}^2 = \sum_i
\size{\braket{e_i}{\hat{e}_{x,j}}}^2$. So,
\begin{eqnarray*}
\sum_x \norm{P_x\ket{\phi_x}}^2 & \le & 
    \sum_i \sum_{x,j} \size{\braket{e_i}{\hat{e}_{x,j}}}^2 \\
& \le & \sum_i \norm{e_i}^2 \\
& \le &  2^m,
\end{eqnarray*}
since the orthonormal basis~$\{\ket{e_i}\}$ for~$E$ has size at most~$2^m$,
which is a bound on the dimension of~$E$.
\end{proof}

By~(\ref{eqn-starstar}),
the probability of correct decoding is at most~$\sum_x p_x
\norm{P_x\ket{\phi_x}}^2$.
From the claim above, this expression is equal to~$\sum_x p_x \lambda_x$,
where~$0 \le \lambda_x \le 1$ and~$\sum_x \lambda_x \le 2^m$. The
maximum over all such~$\{\lambda_x\}$ of this quantity may easily be
seen to be bounded by the sum of the~$2^m$ largest probability
masses~$p_x$, i.e., by~$P(X,2^m)$. Moreover, for any given~$X$ and~$m$, there
is a natural pair of encoding and decoding functions that achieves this bound.
This shows that the bound is tight.

The above bound on decoding probability can give us sharper bounds
on the number of qubits used in an encoding
than an application of Holevo's theorem. We illustrate this with
an example encoding of~$n$-bits into~$n+1$ orthogonal states~$\ket{i}$.
Half the strings are encoded as~$\ket{0}$, a fourth as~$\ket{1}$, an
eighth as~$\ket{2}$, and so on. A random codeword from this code
can be decoded with
probability exactly~$(n+1)2^{-n}$, which yields the correct answer for
the number of qubits used by invoking our bound.
On the other hand, the mutual information~$I(X\!:\!Y)$
between the encoded string and its decoding is 
$$
{1\over 2} + {2\over {2^2}} + {3\over {2^3}} + \cdots
+ {n\over {2^n}} + {n\over {2^n}},
$$
which sums up to~$2-2^{-(n-1)}$.
This gives us a lower bound of at most~$2$ when combined with Holevo's
theorem.

Note that Theorem~\ref{thm-alt-holevo} may be applied in a communication
complexity context as well, when combined with the lemma due
to~\cite{yao,kremer} mentioned in Section~\ref{sec-rac}. This implies
that if after the exchange of~$m$ quantum bits,~$n$ classical bits are
transferred with success probability at least~$\delta > 0$, then~$m \ge
n - \log{1 \over \delta}$. An application of Holevo's theorem along with
Fano's inequality~\cite{ct} would result in the bound~$m \ge \delta n -
H(\delta)$. This lower bound is a crucial ingredient in
proving the quantum communication complexity of the inner product
function~\cite{kremer}. Our result gives a bound similar to that
shown in~\cite{astvw} for computing Inner Product,
but does not seem
to generalize to the case of entanglement assisted communication
considered in~\cite{cdnt}.

\subsection*{Acknowledgements}

I would like to thank Dorit Aharonov for suggesting the
possibility of enhanced models of quantum finite automata,
which motivated much of this work, Andris Ambainis for useful
comments on the paper, Umesh Vazirani for many discussions that
lead to crucial insights and for help with the presentation of the
results, and the referees for their feedback on the paper.

\end{document}